# A Data-Fusion-Assisted Telemetry Layer for Autonomous Optical Networks

Xiaomin Liu, Huazhi Lun, Ruoxuan Gao, Meng Cai, Lilin Yi, Weisheng Hu and Qunbi Zhuge

*Abstract*— For further improving the capacity and reliability of optical networks, a closed-loop autonomous architecture is preferred. Considering a large number of optical components in an optical network and many digital signal processing modules in each optical transceiver, massive real-time data can be collected. However, for a traditional monitoring structure, collecting, storing and processing a large size of data are challenging tasks. Moreover, strong correlations and similarities between data from different sources and regions are not properly considered, which may limit function extension and accuracy improvement. To address abovementioned issues, a data-fusion-assisted telemetry layer between the physical layer and control layer is proposed in this paper. The data fusion methodologies are elaborated on three different levels: *Source Level*, *Space Level* and *Model Level*. For each level, various data fusion algorithms are introduced and relevant works are reviewed. In addition, proof-of-concept use cases for each level are provided through simulations, where the benefits of the data-fusion-assisted telemetry layer are shown.

## I. INTRODUCTION

IN the past decade, optical transmission and networking technologies have been rapidly evolving. In particular, the advance of digital signal processing (DSP) based coherent transceiver enables to approach the capacity limit of a classical fiber channel, and it has become relatively difficult to further increase network capacity through innovations in transmission technologies. Such a situation provides a strong incentive to build a more intelligent optical network for further reduction of cost per bit by more efficiently utilizing network resources. In the meantime, the development of coherent technologies and flex-grid reconfigurable optical add-drop multiplexer (ROADM) has driven the deployment of elastic optical networks (EON) [1], which support flexible modulation formats, channel spacing and throughput. Furthermore, software-defined networks (SDN) technologies [2]–[4] provide a centralized architecture to control, optimize and reconfigure physical layer optical networks, flexibly .

The ultimate goal is to design a closed-loop autonomous optical network, which maximizes system capacity and minimizes operational expenditure. Typical optical network applications include network planning [5], [6], physical layer optimization [7], [8], resource allocation [9]–[11], failure management [12], [13], and so forth. To support the closed-loop automation design of these applications, real-time data collection through telemetry is required [14]. In long-haul optical communications, a single link consists of many elements such as optical transceiver, optical amplifier and ROADM. Transmitted signals are affected by various impairments including transceiver noise, fiber linear effects, and fiber nonlinear effects. Moreover, since tens of channels share one fiber link in wavelength-division multiplexing (WDM) systems, there exists channel crosstalk through fiber nonlinearities. Therefore, real-time data collection is a complex task in optical networks. Recently, artificial intelligence (AI) technology is regarded as a promising solution to achieve autonomous optical networks, but it significantly enhances the requirement for telemetry capabilities.

Currently, the research on optical telemetry, also known as optical performance monitoring (OPM), is usually focused on single function design. For example, optical signal-to-noise ratio (OSNR) monitoring has been widely investigated based on an optical spectrum analyzer (OSA) or a coherent receiver [15]–[19]. From a network perspective, other issues regarding telemetry have not been discussed. First, since there are a large number of monitoring elements in a network and the speed of optical communication is extremely high, massive telemetry data would be generated, leading to a problem in storage space. Second, strong correlations between data sources exist, including different elements across multiple link elements in one channel and between channels. They can be exploited to develop more telemetry functionalities and increase their accuracy. Therefore, the design of telemetry from a network perspective will become an important topic in the context of AI-driven autonomous optical network.

In [20], we review and discuss optical network applications and the required monitoring functionalities. In this paper, a telemetry layer is proposed to address the abovementioned issues and support upper-layer applications. With the aid of the telemetry layer, data from different sources can be processed

Manuscript received XXX XX, 2020; revised XXX XX, 2020; accepted XXX XX, 2020. Date of publication XXX XX, 2020; date of current version XXX XX, 2020. This work was supported by National Natural Science Foundation of China (Grant No. 61801291), Shanghai Rising-Star Program (Grant No. 19QA1404600), and National Key R&D Program of China (Grant No. 2018YFB1801200). (*Corresponding author: Qunbi Zhuge*.)

The authors are with the State Key Laboratory of Advanced Optical Communication Systems and Networks, Shanghai Institute for Advanced Communication and Data Science, Shanghai Jiao Tong University, Shanghai 200240, China (e-mail: xiaomin.liu@sjtu.edu.cn; huazhi.lun@sjtu.edu.cn; echo.xuan@sjtu.edu.cn; caimeng0922@sjtu.edu.cn; lilinyi@sjtu.edu.cn; wshu@sjtu.edu.cn; qunbi.zhuge@sjtu.edu.cn)

before being sent to the control layer, which alleviates the communication and storage burden. Moreover, in the telemetry layer, massive data generated in various domains and regions forms a flexible repository which breaks the barriers between various equipment and regions to achieve precision improvement and function extension. In this layer, data fusion [21] algorithms are the core techniques to preprocess the massive data collected from optical networks. Among these algorithms, AI-based methods are the most important one. Based on data fusion, a large amount of data from multiple heterogenous sources can be aggregated to build an information-integration monitoring system.

We firstly provide an overview of the architecture and functionalities of the proposed telemetry layer. The data fusion methodologies in the telemetry layer are elaborated based on three levels of telemetry perspective: 1) *Source Level,* 2) *Space Level,* 3) *Model Level.* For each level, we describe and analyze the useful algorithms for fusing information from the physical layer and also review many relevant monitoring algorithms. To better demonstrate the function and superiority of the proposed telemetry layer, we also provide use cases in each level to prove the concept.

The rest of this paper is organized as follows. In Section II, we elaborate the architecture of the telemetry layer and briefly introduce the data fusion methods on different levels. In Section III to V, algorithms for fusing data on the abovementioned levels in the telemetry layer are introduced. Moreover, for each section, we demonstrate the application data fusion algorithms in some use cases. In the Section VI, conclusions are obtained by summarizing the advantages of applying the data-fusion-assisted telemetry layer in autonomous optical networks.

## II. THE ARCHITECTURE OF THE TELEMETRY LAYER AND THE DATA FUSION METHODS ON DIFFERENT LEVELS

As shown in Fig. 1, between the control layer and the physical layer, a telemetry layer is proposed to process and fuse data from multiple heterogenous sources. After fusing information, the monitoring accuracy can be improved. In addition, only a small number of monitoring results need to be transferred to the control layer, which can significantly reduce the storage space. Moreover, by processing the extracted data properly, more monitoring functions can be developed to support upper-layer applications.

As is mentioned earlier, to efficiently extract useful information from various data in the physical layer, data fusion algorithms are the core techniques. Different algorithms view data from different perspectives, thus achieving various monitoring goals flexibly. To describe these data fusion algorithms, we categorize them into three levels: 1) *Source Level,* 2) *Space Level,* 3) *Model Level*. The different levels of data fusion methodologies are also depicted in Fig. 1 and elaborated as follows.

*Source level*: On the source level, information from heterogenous sources are fused. In optical transmission, there are two typical sources to extract information: the optical domain and digital domain. In the optical domain, since signals transmit over a long distance, various optical components involve, such as fiber, lumped Erbium-doped fiber amplifiers (EDFA) and wavelength selective switches (WSS). Many impairments are caused by these link elements and related information can be obtained for link monitoring. In the digital domain, many DSP algorithms are utilized in a coherent transceiver to compensate for different transmission impairments, which also provide opportunities to extract impairment-related information for link monitoring. Since the impairments occur during transmission and are subsequently compensated at the receiver, there are strong correlations between the information from the digital domain and the optical domain. Traditional methods usually only consider information from one specific domain. However, by combining information from the two different sources in the telemetry layer, more accurate monitoring is hopeful to be achieved.

*Space Level*: On the space level, data from one region or multiple regions can be fused by different methods. For data in one region, in a WDM system, signals from different channels share the same fiber and go through the same equipment. They experience identical or similar impairments so that the strong correlations between each other exist. Therefore, fusing information from relevant channels can provide stronger monitoring capabilities. For multiple regions, data with different uncertainty distributions can be employed to improve the robustness and generalization ability of the global applications for all network regions.

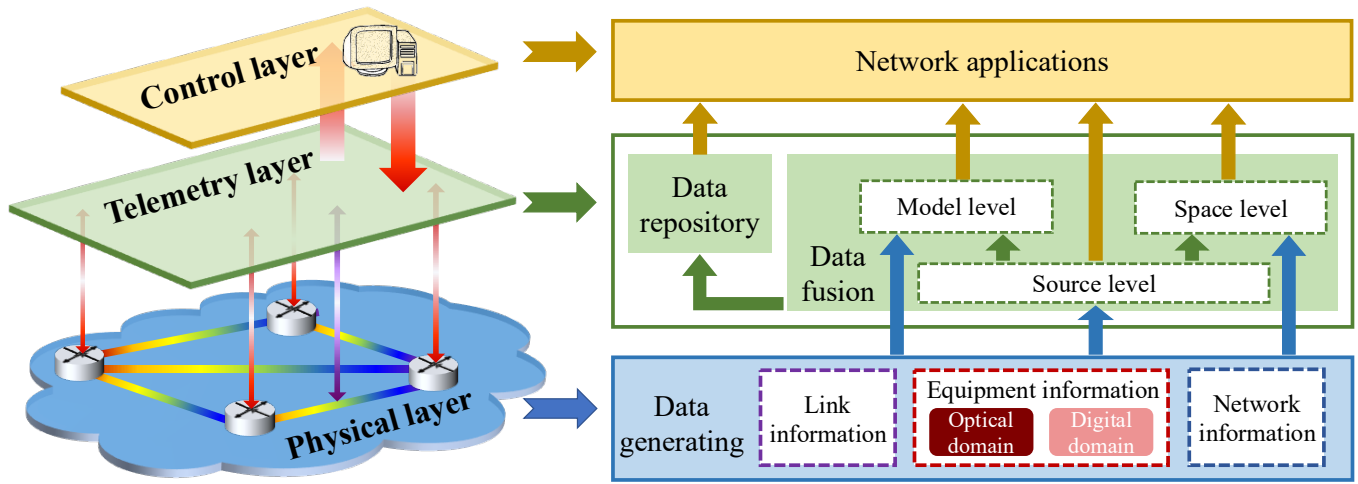

Fig. 1. The framework of the data fusion methodologies for intelligent network telemetry.

*Model Level*: On the model level, models with homogenous or heterogenous forms can be aggregated to combine their advantages and make up for their shortcomings. For heterogeneous models, i.e. analytical models and machine-learning-based (ML) models, they can be ensembled to perform a better estimation since they provide information from diverse perspectives. For homogeneous models, i.e. ML-based models, they can be ensembled to transform many weak learners to become a strong learner.

These three levels of data fusion algorithms can be applied parallelly, hierarchically or individually. By fusing data in the telemetry layer, more monitoring functions can be provided and the monitoring accuracy can be improved. In the following sections, we will elaborate the data fusion methods in the telemetry layer on each level and provide use cases to prove the concept.

### III. Data-Fusion-Assisted Telemetry on the Source Level

#### A. Monitoring with Data from Heterogenous Sources

In optical networks, fusing information from the optical domain and digital domain can help to improve link monitoring. In the optical domain, a number of parameters reflecting system status can be obtained. For a fiber span, fiber attenuation can be monitored by measuring the power at the both ends of the span. For an amplifier, the average launch power of signals at the output of the amplifier can be measured. If an OSA is available, the power of each channel can be further obtained. For a WSS, the optical spectrum before and after each WSS can be also measured by the OSA. The information that can be obtained in the optical domain are summarized in Table I. With the monitored information, we can gain insights into the system status. For example, in [7], the nonlinear impairments caused by the Kerr effect are estimated by analyzing the optical spectrum. In [22], the gain fluctuations of the EDFA on the optical link are obtained with the help of the optical spectrum. In [12], the cause of the soft failure between filter shift (FS) and filter tightening (FT) is identified by analyzing the optical spectrum.

TABLE I
INFORMATION IN OPTICAL DOMAIN

| Link components | Relevant information |
| --- | --- |
| Fiber | Fiber attenuation |
| EDFA | Amplification gain<br>Average launch power<br>Per channel signal power |
| WSS | Optical spectrum<br>Per channel signal power |

Besides the optical domain, the deployed coherent receiver can be easily extended to provide monitoring functions. The typical DSP consists of chromatic dispersion compensation (CDC), adaptive equalizer, carrier recovery (CR), nonlinearity compensation (NLC), clock recovery and so on. From each DSP module, different end-to-end information of the link can be monitored, and they are summarized in Table II. With the help of these DSP information, the status of the system can be estimated and optimized. In [23], the tap of the adaptive filter is used to estimate the channel response to perform pre-compensation of the filtering effect of WSS. In [24], the filter tap is used for soft failure detection. In [25], the receiver bit-error-ratio (BER) is monitored, and the nonlinear impairment of fiber is estimated by analyzing the BER histogram. In [26], the amplitude noise correlation, the phase noise correlation and the chromatic dispersion are extracted from the DSP to estimate the nonlinear noise. In [27], the phase noise is extracted from the CR module to estimate the nonlinear noise of the system.

TABLE II
INFORMATION IN DIGITAL DOMAIN

| DSP module | Available information |
| --- | --- |
| CDC | Digital spectrum<br>Accumulated chromatic dispersion |
| Adaptive equalizer | filtering effect<br>Transceiver I/Q error<br>Polarization effects |
| CR | Laser phase noise<br>Fiber nonlinear phase noise<br>Frequency offset |
| NLC | Nonlinear noise |
| Clock recovery | Clock mismatch |

#### B. Methodologies for Fusing Data on the Source Level

From the description above, both the optical domain and digital domain can provide useful information about the optical communication system. For long-haul transmissions, major impairments occur in the optical domain and are compensated in the digital domain. Therefore, the information from the two domains can have strong correlations. For many previous researches, only information from one specific domain is employed, which limits the monitoring accuracy to some extent. By fusing data from different domains, a higher accuracy can be achieved and more monitoring functions can be potentially developed.

The data obtained from the physical layer can be classified into three categories [21]. 1) *Complementary*: the obtained data is from different sources but reflects the same problem, which can be aggregated to provide a more complete picture for the monitoring task. For example, the signal spectrum from the CDC and adaptive filter reflects the channel response in a complementary way, which can help to monitor the filtering effect with a higher accuracy. 2) *Redundant*: when two or more input sources provide information about the same target and could thus be fused to increase the confidence. For example, both the optical spectrum obtained from an OSA and the digital spectrum extracted from a coherent receiver reflect the same channel response. Therefore, combining spectrums from the two sources can improve the monitoring of the filtering effect. 3) *Cooperative*: when information is from different sources and acquired by different algorithms, they can be integrated to assemble the knowledge and improve the performance of each individual. For example, in [28], results estimated based on the Gaussian noise (GN) model and monitored information from the physical layer are combined to monitor the self-phase modulation (SPM) more accurately. For different cases, a proper fusion method is needed to combine the knowledge for better link monitoring. The overall framework to fuse data from heterogenous sources is shown in Fig. 2 and specific data fusion methods on the *Source Level* are elaborated as follows.

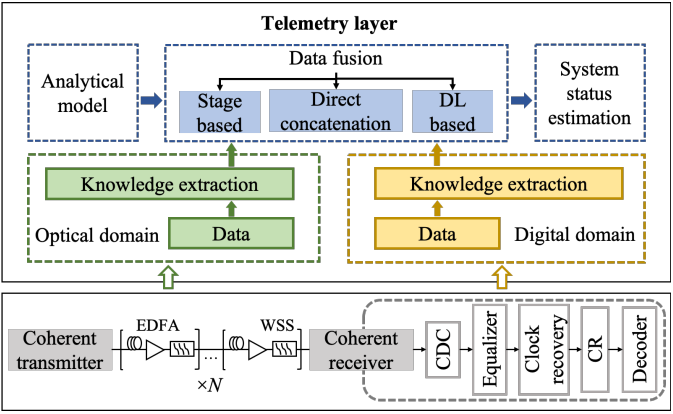

Fig. 2. The overall framework for source level data fusion approaches on the telemetry layer.

**Direct-concatenation-based data fusion**: The easiest way to fuse data is the direct concatenation [29]. This method treats the knowledge from different sources equally and combine them into a feature vector sequentially. Afterwards, these features are input to a function or a ML model. In [30], authors concatenate the features extracted from optical spectrum and the tap value from the equalizer of a coherent receiver to localize the abnormal WSS in a link. This method is simple to process and is useful when the number of features is not large. However, disadvantages exist [10]. First, overfitting can be easily caused when the dataset is small. Second, the internal nonlinear relationship between features from different sources is ignored, which limits the monitor to extract more complex knowledge such as the shape of the spectrum. Third, it may cause high redundancy since data may have correlations between each other, which can result in a large computational load.

**Stage-based data fusion**: This method decomposes a problem into different stages, and information from different domains is used accordingly. At different stages, the algorithms used for knowledge extraction may be different, making it possible to deploy the data fusion framework into different layers. The stage using simple algorithms can be integrated into the receiver, and the stage requiring more computation resources can be deployed into the telemetry layer. In this way, the communication burden can be alleviated to some extent. Another advantage is that at different stages, the information obtained from different sources can be analyzed by the most proper algorithm according to their respective characteristics. In [31], a two-stage soft failure identification scheme has been proposed.

**Deep-learning-based data fusion**: The last one is the deep learning (DL) based data fusion method. With the fast advance of DL technologies, its powerful feature extraction capability makes it a new way for data fusion. With various supervised, unsupervised algorithms, DL can learn multiple levels of representation and abstraction of data. In addition to outputting results directly, deep learning can also learn new feature representations and these middle features are then input to another predictor (typically a shallow artificial neural network). These new feature representations are proved to be more useful than hand-crafted features [32].

*C. Use Case for Source Level Data-Fusion-Assisted Telemetry*

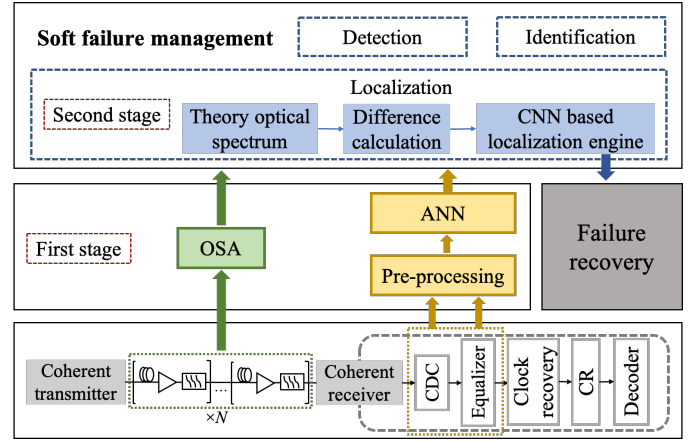

Fig. 3. Diagram of the two-stage-based soft failure identification and localization.

In this section, we describe a use case using the stage-based data fusion scheme on the *Source Level*. We extend our previously proposed one-stage structure in [30] to a two-stage structure which can detect and localize soft failures at the same time. At the first stage, the monitoring features include the average value, maximum value, minimum value, standard deviation of the equalizer tap, the centroid and the 3-dB bandwidth of the digital spectrum. These features are analyzed by an artificial neural network (ANN) and a coarse localization result is output. Since the first stage gives an approximate location of the soft failure, there is no need to upload every optical spectrum of an optical link after each WSS. Therefore, at the second stage, only optical spectrums around the approximate location are needed, which significantly reduces the communication burden. Afterwards, the difference between the theoretical optical spectrum and the monitored spectrum is calculated. The theoretical optical spectrum $s(f)$ at the frequency $f$ is calculated by

$$s(f) = \frac{1}{2}\sigma\sqrt{2\pi}\left[\text{erf}\left(\frac{\frac{B}{2}-f}{\sqrt{2}\sigma}\right) - \text{erf}\left(\frac{-\frac{B}{2}-f}{\sqrt{2}\sigma}\right)\right] \quad (1)$$

$$\sigma = \frac{BW_{otf}}{2\sqrt{2ln2}} \quad (2)$$

where $B$ and $W_{otf}$ are the bandwidth of the WSS filter and the steepness of the filter edge, respectively. Then, the difference is input into a convolutional neural network for the accurate soft failure identification and localization.

For the deployment of the algorithm, the first stage is just a shallow hidden neural network that can be deployed in each receiver. The second stage needs a convolutional neural network, which can be deployed in the central telemetry layer. Afterwards, the failure locations are uploaded to the control layer. The whole process described above is shown in Fig. 3.

To validate the proposed method, we performed simulations and the simulation setup is shown in Fig. 4.

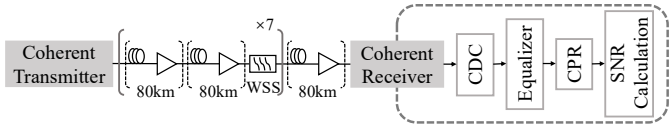

Fig. 4. The simulation setup.

At the transmitter side, the transmitted data is first generated and mapped into 16-QAM symbols. The symbol rate is set to 35 Gbaud. After up-sampling and pulse shaping by a raised-root-cosine (RRC) filter with a roll-off factor of 0.02, the signal is launched into the fiber. The fiber link consists of standard single mode fiber (SSMF) with a fixed span length of 80 km. The attenuation of fiber is 0.2 dB/km, the dispersion coefficient is 16.7 ps/nm/km. The nonlinear Kerr effect is ignored in our simulation. After each span, an EDFA is used to completely compensate the span loss. The noise figure (NF) of each EDFA is set to 5 dB. The total length of the link is 1200 km, and after every two spans a WSS is inserted.

At the receiver side, a CDC module is first applied to the signal in frequency domain. After that, the matched filter is applied followed by an adaptive equalizer based on the principle of the least-mean-square (LMS) criterion. Finally, the CPR is applied and then the signal-to-noise ratio (SNR) can be obtained. The equalizer tap values are extracted. To simulate the FS, we sweep the center frequency of each WSS from 15 GHz to 20 GHz. To simulate the FT, we sweep the bandwidth of each WSS from 20 GHz to 25 GHz. Finally, we generate in total 1402 examples. 70% of them are for training, and 30% of them are for testing.

At the first stage, we use a 2 hidden-layer ANN. The results are shown in Fig. 5 and Fig. 6. Although the accuracy reaches 93%, some samples are still misclassified, and we find that all the misclassifications are the neighboring ones of the correct one. This is because the end-to-end DSP information cannot provide sufficient information when the soft failure locations are close.

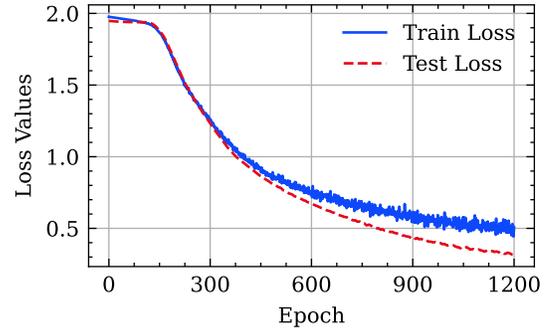

Fig. 5. The loss curves of the ANN at the first stage.

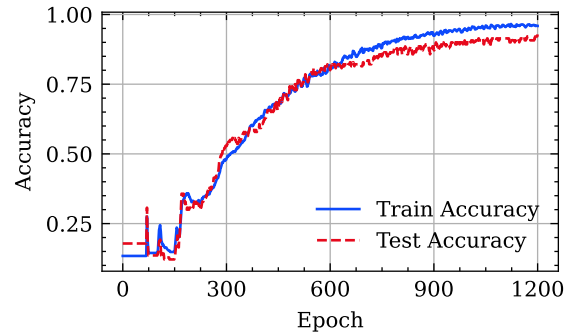

Fig. 6. The accuracy of the ANN at the first stage.

To further improve the accuracy, the second stage collects the optical spectrums around the soft failure location determined by the first stage. The difference between the theoretical optical spectrum and the actual one is analyzed by a convolutional neural network (CNN) and the soft failure cause is identified and localized. The accuracy improves from 93% to 100% compared with the first stage output, which is reasonable according to many previous researches in the field of computer vision [12]. Therefore, in the telemetry layer, by the stage-based data fusion method, the optical spectrum and the DSP information are combined to identify and localize the soft failure accurately.

IV. DATA-FUSION-ASSISTED TELEMETRY ON THE SPACE LEVEL

A large number of optical components are employed to support communication requirements between various nodes, building an optical network in each region. Considering the scale of the signal transmission, in different regions on the *Space Level*, different optical networks exist. In this section, we introduce data fusion methodologies on the *Space Level* from two perspectives: 1) fusing data from one region; 2) fusing data from multiple regions.

*A. Fusing Data from One Region*

For an optical network in one region, physical layer impairments tend to have strong correlations and high similarities, which provides a suitable scenario for data fusion. For WDM transmissions in one link, signals share the same optical components, therefore inducing similar or relevant impairments and noises. In the same fiber, signals induce cross phase modulation impairments for each other. For the same EDFAs, the NF spectrums are continuous functions and show gain/loss wavelength dependency. For the same WSS, when failures occur, all signals may be influenced. Therefore, utilizing data from signals in one link can provide various relevant and similar information to improve link monitoring.

For signals transmitted in different links in one region, they also have some correlations and similarities. Firstly, many signals in different links may share the same optical multiplex section (OMS) with the same optical equipment in the network. Secondly, in one region, optical components are usually from the same vendors with the same configuration settings. Therefore, even though these signals may not share the same link or equipment, they may experience impairments with similar characteristics.

Considering the strong correlations and similarities of the optical channels in one region, data from each channel can be fused to improve link monitoring. However, the similarity is difficult to assess directly. This is because the configuration parameters of each channel are generated to be a high-dimensional vector. The importance and weight of each parameter are unknown so that it is difficult to choose an appropriate function to measure the similarity. To solve this, using ML-based methods, metric learning [33] can help to learn the relationship of each signal in a supervised manner. The similarities of each signal and configuration setting are learned iteratively so that they can be employed to make estimations when new signals are to be established or aggregated for achieving more accurate monitoring. Moreover, traditional algorithms such as K-Nearest Neighbor (K-NN) or linear regression are also useful tools for extracting the correlations between signals. For example, in [34], when many signals share the same EDFAs, Gaussian process regression is applied to estimate the OSNR in each channel without any prior knowledge. In [35], a method evaluating the similarity of each signal is proposed to build a cognitive optical network. To further explore information from relevant signals, in [36], a graph neural network is designed to consider the interference between each node and each fiber for failure localization. To optimize estimation performance, active learning can also be utilized to select the most useful information from a large number of signals [37].

*B. Fusing Data in Multiple Regions*

In different regions, signals also have some similarities since they all utilize optical elements with the same functions, such as fiber, EDFA and WSS. To make full use of this kind of similarity, we can fuse these data and further improve the robustness and generalization ability of the global models for various applications.

As shown in Fig. 7, model assigned to each region cannot perfectly match the practical system at the beginning and thus an adaptation is needed. To improve the performance of the application in the new region, the global model also needs to be improved. In this case, data from multiple regions can be aggregated to adjust the global model. However, a large amount of data is transferred from each region, which can cause a large communication burden if simply sending data to the control plane without any processing for conventional retraining or updating. To solve this, instead of uploading data in all regions, models updated with local data in each region can be applied to conduct data fusion, which decouples fusing process from the direct access to the physical-layer data [38]. This kind of data fusion method is called *federated learning* [38], which is a decentralized learning framework and useful for NN-based algorithms. Proposed in [39], the distributed and centralized learning framework has been analyzed to build a learning life cycle. In [40], the centralized and distributed processing are cooperated to build an SDN-based hybrid framework for transmission performance estimation.

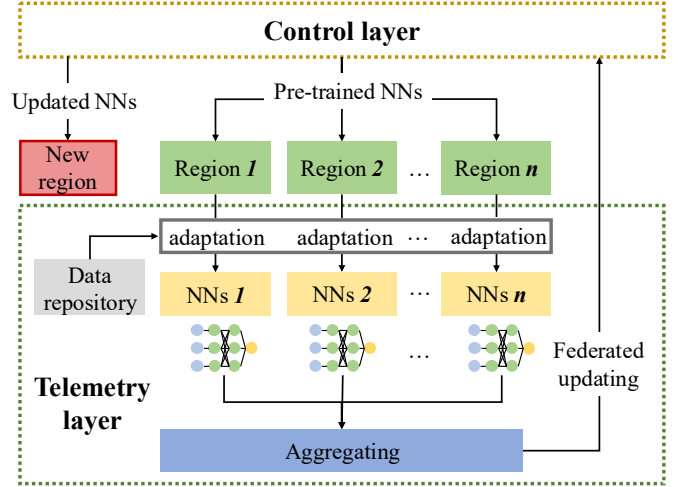

Fig. 7. The structure of the federated-learning-based data fusion framework on the telemetry layer.

*C. Use Case for Space Level Data-Fusion-Assisted Telemetry*

Here we apply the federated-learning-based data fusion to improve the robustness and generalization ability of the global model for nonlinear noise estimation. Based on neural networks, nonlinear noise can be estimated [41]. However, since the parameter uncertainty, such as the input power, caused by inaccurate measurements existing in practical systems, the pre-trained model needs to be customized to be compatible with different regions [42]. As shown in Fig. 7, with the aid of the telemetry layer, data extracted from physical layer can be easily obtained by local applications for adaptation. Afterwards, these locally adjusted models can be fused to update the global model. Therefore, the robustness and generalization ability of the global model can be further improved.

The simulation system is fixed-grid with a symbol rate of 35 Gbaud and a channel spacing of 50 GHz. The symbol length for each example is $2^{16}$. At the transmitter (Tx) side, RRC pulse shaping with a roll-off of 0.02 is applied. EDFAs are employed after each fiber span. During transmission, the fiber nonlinearity is simulated by SSFM. At the receiver (Rx) side, the center channel is filtered out for the signal processing, which includes CDC, matched filtering, down-sampling and phase de-rotation to remove the phase rotation caused by the nonlinearity. Finally, the nonlinear SNR is obtained. We generate 1000 examples in this simulation system. The configurations of links and signals are randomly chosen from

Table III. For each example, some of the channels are randomly set to be idle. The fiber types include SSMF, enhanced large effective area fiber (ELEAF), and pure silica core fiber (PSCF).

TABLE III
SUMMARY OF LINK CONFIGURATIONS AND PARAMETERS

| Fiber type | SSMF, ELEAF, PSCF |
|---|---|
| Span length (km) | 80 |
| Span number | 3:1:21 |
| Launch power (dBm) | -3:1:3 |
| Modulation format | DP-QPSK, DP-16QAM |
| Channel number | 3:1:21 |
| Symbol rate (Gbaud) | 35 |

After generating data from simulations, we build a two-hidden-layer neural network based on Pytorch. The first and second layers both have 32 nodes. The activation function of each layer is LeakyReLU. The parameters of the neural networks are updated following the gradient descent. The input parameters of the neural networks are shown in Table IV.

TABLE IV
INPUT FEATURES OF THE NEURAL NETWORKS

| | |
|---|---|
| | Span number |
| | Span length |
| | Launch power |
| Input features | Link length |
| | Net chromatic dispersion |
| | Average gamma of fiber spans |
| | Average alpha of fiber spans |
| | Number of WDM channels |

TABLE V
THE MEAN AND VARIANCE OF THE INPUT POWER UNCERTAINTY FOR FOUR ESTABLISHED REGIONS

| Region | 1 | 2 | 3 | 4 |
|---|---|---|---|---|
| Mean (dB) | 0.3 | 0.4 | 0.6 | 0.5 |
| Variance (dB$^2$) | 0.5 | 0.3 | 0.2 | 0.3 |

We assume that the global model has been assigned to four regions with different input power uncertainties. The uncertainty distribution follows a Gaussian distribution and the mean and variance of the uncertainty for each region are shown in Table V. To customize the global model, each region is assumed to be able to provide 200 examples with different link/signal configurations in our simulations. In the telemetry layer, the real nonlinear SNR of these examples in practical systems can be monitored by the method proposed in [26] and employed for adaptation. After obtaining the customized models, they are aggregated to update the global model in the control plane. Here we apply the Fed-SGD method [38], of which the parameters of the neural networks are updated by

$$f(\omega) = \sum_{k=1}^{K} \frac{n_k}{n} F_k(\omega) \quad (3)$$

where $f(\omega)$ and $F_k(\omega)$ denote the parameters of the model after federated updating and the parameters of the model in the $k$-th region, respectively. Assuming there are $K$ regions in total, the $n_k$ and $n$ are the number of examples in the $k$-th region and all regions, respectively. In our simulations, four customized models from the regions where the global model have been assigned to are applied for updating with the Fed-SGD.

To evaluate the performance of the federated-learning-based data fusion, we apply the updated global model to a new region with 200 examples. In this region, we generate 2000 kinds of different input power uncertainties which all follow the Gaussian distribution with a mean ranging from 0 dB to 1 dB and a variance from 0 dB$^2$ to 0.4 dB$^2$. The mean and variance in each case are set randomly. Afterwards, the estimated mean square errors (MSE) of these 2000 cases are calculated. The results are shown in Fig. 8.

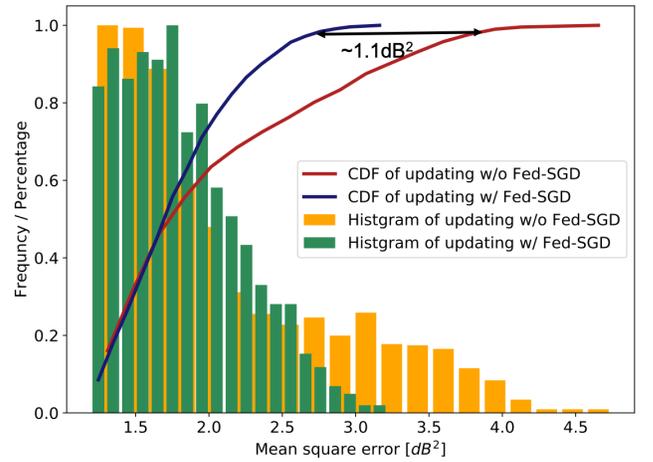

Fig. 8. The error histogram and the cumulative distribution function of the MSE of 2000 cases with different uncertainty distributions.

TABLE VI
THE MEAN AND VARIANCE OF THE INPUT POWER UNCERTAINTY FOR FIVE SPECIFIC CASES

| Case name | 1 | 2 | 3 | 4 | 5 |
|---|---|---|---|---|---|
| Mean (dB) | 0.5 | 0.3 | 0.5 | 0.6 | 0.6 |
| Variance (dB$^2$) | 0.5 | 0.3 | 0.4 | 0.1 | 0.3 |

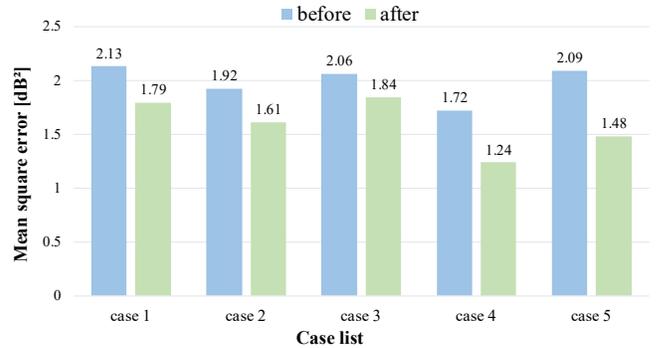

Fig. 9. The comparisons for the MSE of each case before and after federated-learning-based data fusion.

As shown in Fig. 8, after the federated updating, the accuracy of the model employed to the new region has improved. The deviation including 95% cases reduces from 3.8 dB$^2$ to 2.7 dB$^2$, which means the global model after federated learning has a better robustness and generalization ability. Therefore, when the customization procedure cannot be applied, the global model after updating can perform a better estimation. We select five kinds of uncertainty distributions specifically to show the performance of the global model before and after the federated updating. The uncertainty distributions of these five cases are shown in Table VI. The results of the estimation performance are shown in Fig. 9. The results of these five cases show that after updating the global model with the Fed-SGD, the MSE of the estimation results significantly reduce, which reveals that the robustness and generalization ability of the global model has been improved by fusing information from multiple regions.

## V. Data-fusion-Assisted Telemetry on the Model Level

To estimate various impairments in the physical layer, many models have been proposed, which can be divided into two types: 1) analytical model and 2) data-driven model. The former is generally not able to guarantee accuracy for all scenarios, and the latter usually requires a large amount of data to build. In the telemetry layer, models can be aggregated to combine their advantages for a higher accuracy. Next, we introduce two types of model fusion methods on the telemetry layer.

### A. Fusing Heterogenous Models

In optical transmission, various analytical models, such as GN model [43] and SSFM [44], can provide rigorous theoretical estimations for link impairment. However, in certain cases the accuracy is limited. Based on analytical models, data-driven models only need to fit the residuals of the theoretical results rather than the real relationship with a high-dimensional non-convex loss function. As shown in Fig. 10, the results of analytical models can speed up the convergence during training and help algorithms avoid being trapped in sub-optimal hypotheses [45]. We were inspired by the idea of residual learning framework [46] to combine the training-based monitoring method with the GN model for further improving nonlinear noise monitoring [41].

### B. Fusing Homogeneous Models

The second type of fusion methods is to combine many homogeneous *weak leaners* to build a *stronger learner* [47], which is also called model ensembles. This kind of model fusion are often applied when ML-based models are employed.

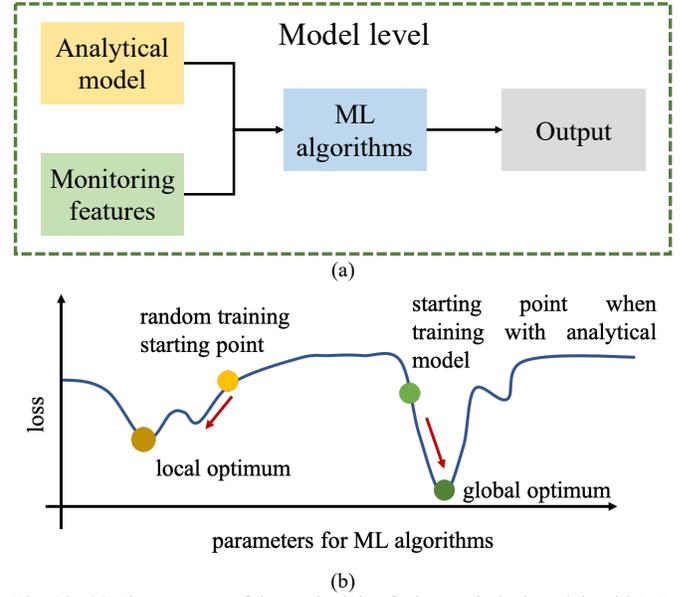

Fig. 10. (a) The structure of the method that fusing analytical models with ML algorithms. (b) The training process of the ML algorithms with or without analytical models.

In this way, the accuracy the applied ML-based methods can be further improved. There are four kinds of typical ensemble algorithms: *Boosting* [48]*, Bagging* [49]*, Blending* [50] and *Stacking* [51]. The brief descriptions of these four methods are shown in Fig. 11 and they are elaborated as follows.

*Boosting* methods generate learners sequentially [52] and train them iteratively. For each iteration, the weights of the poorly predicted examples are improved successively for the following training. Among many *Boosting* algorithms, *AdaBoost* [53] and *XGBoost* [54] are largely employed in many cases. Moreover, the *XGBoost*-based method can analyze the importance of the input features, which improves the interpretability of ML models. In [55], the importance output of *XGBoost* is applied to successfully reveal the most significant causes of failures in optical networks. However, since *Boosting*-based methods are trained hierarchically, the computational load is very high when the structure of the algorithm is complex.

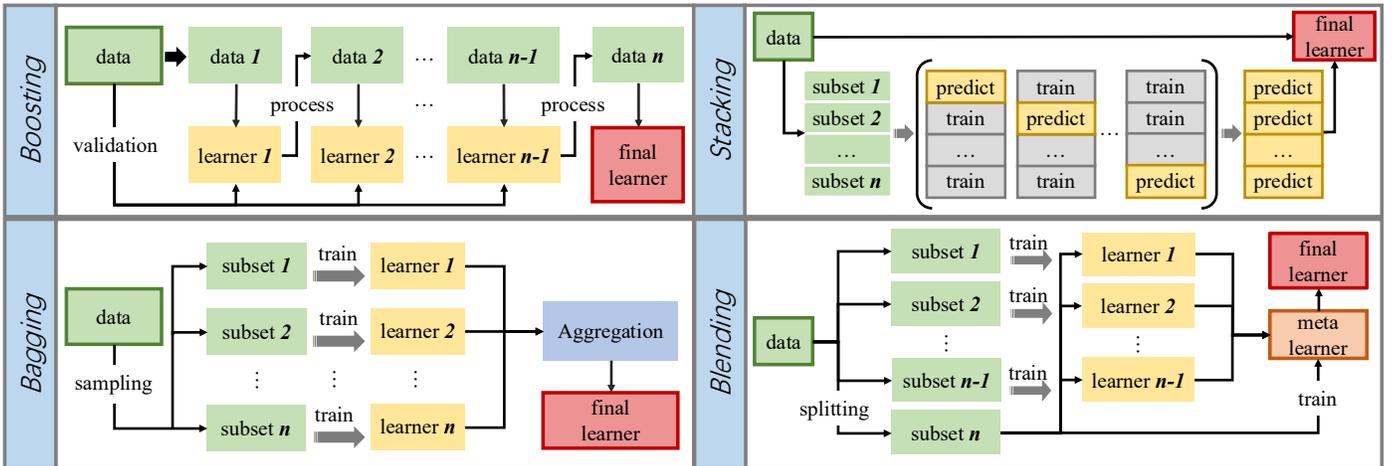

Fig. 11. The structure of the model ensemble methods: *Boosting, Bagging, Stacking, Blending*.

Unlike *Boosting*, bootstrap aggregation, *Bagging* for short, is another ensemble method which can train multiple ML engines parallelly with different sets of data sampled from the training set. Afterwards, estimations of these engines are combined after processing like averaging or voting. *Bagging* can significantly reduce the variance of the model so that it outperforms *Boosting* in situations with noisy data [52], [56]. *Random Forest* is one of the most famous *Bagging* methods, which also shows a good performance in optical monitoring for failure detection [13].

To combine heterogeneous ML engines, *Blending* and *Stacking* use a meta-learner to generate the output of each learner and fuse estimations accordingly. During training, *Blending* simply splits data for each training stage while *Stacking* employs k-folds cross-validation [57]. These two methods can learn the weight and importance of each learner automatically without any manual interference.

### C. Use Case for Model Level Data-Fusion-Assisted Telemetry

In this use case, we combine a ML model with an analytical model to improve monitoring accuracy. Taking nonlinear noise monitoring as an example, as proposed in [41], the neural networks can be combined with the GN model to further improve precision. As shown in Fig. 12, for each transceiver, DSP blocks can provide information related to the nonlinear impairments. With the aid of the neural networks, the nonlinear noise can be monitored [41]. However, purely relying on the monitoring data in the receiver cannot provide very high accuracy. In the telemetry layer, the GN model with the signal and link configurations parameters as the input is combined with the original monitoring features using a neural network to provide a more accurate estimation of nonlinear SNR.

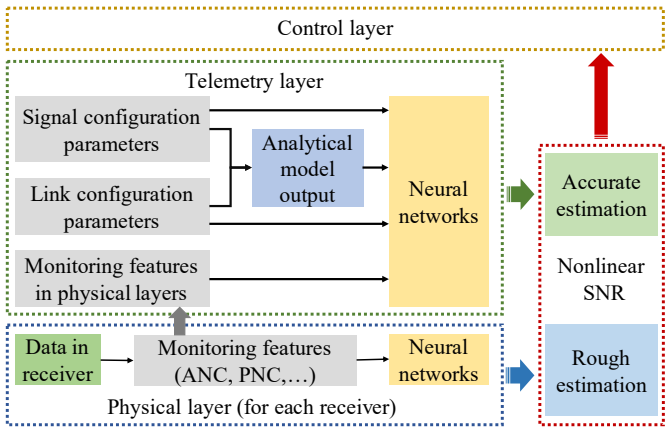

Fig. 12. The monitoring structure with corporations between each layer.

We generate 1000 examples by SSFM, the generating procedure is the same as that in Section IV. The monitoring features are the amplitude noise correlation (ANC) and phase noise correlation proposed (PNC) proposed in [26]. In this simulation, the summation lengths of the ANC and PNC are 3 and 15, respectively. At the receiver side, ANC and PNC are sent to neural networks together with the accumulated chromatic dispersion (CD) and the WDM channel number. We use 750 examples for training and 250 examples for testing.

After training the neural network with one 10-node hidden layer in MATLAB, the estimation results are shown in Fig. 13.

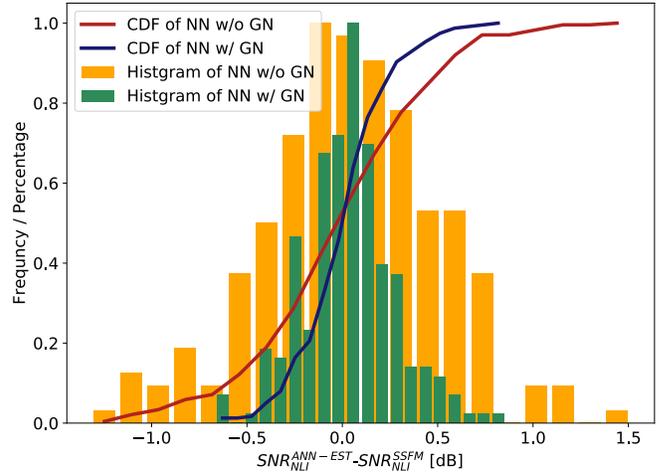

Fig. 13. The error histograms and cumulative distribution functions of the nonlinear SNR estimated by the neural networks in the receiver and in the telemetry layer.

The histogram in Fig. 13 shows that when employing the ANC and PNC, about 95% examples can be estimated with an error less than 1 dB. In the telemetry layer, when combining the GN model, the monitoring accuracy improves. About 95% examples can be estimated with an error less than 0.5 dB. Compared with the estimation made by the neural networks in each receiver, the root-mean-square error (RMSE) of the monitoring result improves from 0.43 dB to 0.23 dB. This is because that in the telemetry layer, more information about the link and signal configurations can be obtained, therefore neural networks can learn more to achieve higher accuracy.

## VI. CONCLUDING REMARKS

A telemetry layer has been proposed to support autonomous optical networks with an information-integration monitoring system. On this layer, information extracted from various link components and receiver digital signal processing blocks can be aggregated, which forms a repository to support upper-layer applications. Based on data fusion, the massive data collected from the physical layer is processed before being sent to the control layer, which alleviates the communication burden and storage burden for network control.

With the aid of the machine learning algorithms, various data fusion methods can be employed on different levels. The approaches for fusing data on the *Source Level, Space Level* and *Model Level* are investigated with some use cases. On the *Source Level*, heterogenous data from digital and optical domains are utilized together to perform a higher estimation precision since different sources can provide correlated information from different aspects. On the *Space Level*, information from shared and/or similar link components are considered in the context of WDM systems and meshed optical networks. On the *Model Level*, multiple models including theoretical and data-driven models are fused to combine their advantages and make up for their shortcomings. With the three levels of data fusion applied parallelly, hierarchically or individually, stronger monitoring capabilities can be built to support closed-loop autonomous optical networks.

REFERENCES

[1] B. C. Chatterjee, N. Sarma, and E. Oki, "Routing and spectrum allocation in elastic optical networks: a tutorial," *IEEE Commun. Surv. Tutor.*, vol. 17, no. 3, pp. 1776–1800, 2015.
[2] M. Channegowda, R. Nejabati, and D. Simeonidou, "Software-defined optical networks technology and infrastructure: enabling software-defined optical network operations [Invited]," *J. Opt. Commun. Netw.*, vol. 5, no. 10, p. A274, Oct. 2013.
[3] F. Hu, Q. Hao, and K. Bao, "A survey on software-defined network and OpenFlow: from concept to implementation," *IEEE Commun. Surv. Tutor.*, vol. 16, no. 4, pp. 2181–2206, 2014.
[4] Y. Li and M. Chen, "Software-defined network function virtualization: a survey," *IEEE Access*, vol. 3, pp. 2542–2553, 2015.
[5] K. Christodoulopoulos, C. Delezoide, N. Sambo, A. Kretsis, I. Sartzetakis, A. Sgambelluri, N. Argyris, G. Kanakis, P. Giardina, G. Bernini, D. Roccato, A. Percelsi, R. Morro, H. Avramopoulos, P. Castoldi, P. Layec, and S. Bigo, "Toward efficient, reliable, and autonomous optical networks: the ORCHESTRA solution [Invited]," *J. Opt. Commun. Netw.*, vol. 11, no. 9, p. C10, Sep. 2019.
[6] I. Sartzetakis, K. Christodoulopoulos, C. P. Tsekrekos, D. Syvridis, and E. Varvarigos, "Estimating QoT of unestablished lightpaths," in *Proc. Opt. Fiber Commun. Conf.*, Anaheim, California, 2016, paper Tu3F.2.
[7] M. Lonardi, P. Ramantanis, P. Jennevé, J. Pesic, A. Ghazisaeidi, and N. Rossi, "Optical nonlinearity monitoring and launched power optimization by artificial neural networks," in *Proc. Eur. Conf. Opt. Commun.*, Dublin, Ireland, 2019, pp. 1-4.
[8] H. Nakashima, Y. Akiyama, T. Hoshida, T. Oyama, T. Tanimura, and J. C. Rasmussen, "Launch power optimization co-operated with digital nonlinear compensator for elastic optical network," in *Proc. Opt. Elect. Commun. Conf.*, Shanghai, China, Jun. 2015, pp. 1–3.
[9] V. Abedifar, M. Furdek, A. Muhammad, M. Eshghi, and L. Wosinska, "Routing, modulation format, spectrum and core allocation in SDM networks based on programmable filterless nodes," in *Proc. Adv. Pho. Conf*, Zurich, 2018, p. NeW3F.4.
[10] S. Behera, A. Deb, G. Das, and B. Mukherjee, "Impairment aware routing, bit loading, and spectrum allocation in elastic optical networks," *J. Light. Technol.*, vol. 37, no. 13, pp. 3009–3020, Jul. 2019.
[11] M. Yang, Q. Wu, and Y. Zhang, "Joint assignment of spatial granularity, routing, modulation, and spectrum in SDM-EONs: minimizing the network CAPEX considering spectrum, WSS, and laser resources," *J. Light. Technol.*, vol. 36, no. 18, pp. 4153–4166, Sep. 2018.
[12] B. Shariati, M. Ruiz, J. Comellas, and L. Velasco, "Learning from the optical spectrum: failure detection and identification," *J. Light. Technol.*, vol. 37, no. 2, pp. 433–440, Jan. 2019.
[13] S. Shahkarami, F. Musumeci, F. Cugini, and M. Tornatore, "Machine-learning-based soft-failure detection and identification in optical networks," in *Proc. Opt. Fiber Commun. Conf.*, San Diego, California, 2018, p. M3A.5.
[14] F. Paolucci, A. Sgambelluri, F. Cugini, and P. Castoldi, "Network telemetry streaming services in SDN-based disaggregated optical networks," *J. Light. Technol.*, vol. 36, no. 15, pp. 3142–3149, Aug. 2018.
[15] T. Tanimura, T. Hoshida, J. C. Rasmussen, M. Suzuki, and H. Morikawa, "OSNR monitoring by deep neural networks trained with asynchronously sampled data," in *Proc. Opt. Commun. Conf.*, Niigata, 2016, pp. 1-3.
[16] L. Xia, J. Zhang, S. Hu, M. Zhu, Y. Song, and K. Qiu, "Transfer learning assisted deep neural network for OSNR estimation," *Opt. Express*, vol. 27, no. 14, p. 19398, Jul. 2019.
[17] F. N. Khan, K. Zhong, X. Zhou, W. H. Al-Arashi, C. Yu, C. Lu, and A. P. T. Lau, "Joint OSNR monitoring and modulation format identification in digital coherent receivers using deep neural networks," *Opt. Express*, vol. 25, no. 15, Art. no. 15, Jul. 2017.
[18] X. Li, L. Zhang, J. Wei, and S. Huang, "Deep neural network based OSNR and availability predictions for multicast light-trees in optical WDM networks," *Opt. Express*, vol. 28, no. 7, p. 10648, Mar. 2020.
[19] A. D. Shiner *et al.*, "Neural network training for OSNR estimation from prototype to product," in *Proc. Eur. Conf. Opt. Commun.*, San Diego, California, 2020, p. M4E.2.
[20] Q. Zhuge, X. Liu, H. Lun, M. Fu, L. Yi, and W. Hu, "DSP-aided telemetry in monitoring linear and nonlinear optical transmission impairments," in *Proc. Opt. Fiber Commun. Conf.*, San Diego, California, 2020, p. M2J.1.
[21] F. Castanedo, "A review of data fusion techniques," *Sci. World J.*, vol. 2013, pp. 1–19, 2013.
[22] A. D'Amico, S. Straullu, A. Nespola, I. Khan, E. London, E. Virgillito, S. Piciaccia, A. Tanzi, G. Galimberti, and V. Curri., "Using machine learning in an open optical line system controller," *J. Opt. Commun. Netw.*, vol. 12, no. 6, p. C1, Jun. 2020.
[23] J. Pan and S. Tibuleac, "Real-time ROADM filtering penalty characterization and generalized precompensation for flexible grid networks," *IEEE Photonics J.*, vol. 9, no. 3, pp. 1–10, Jun. 2017.
[24] S. Varughese, D. Lippiatt, T. Richter, S. Tibuleac, and S. E. Ralph, "Identification of soft failures in optical links using low complexity anomaly detection," in *Proc. Opt. Fiber Commun. Conf.*, San Diego, California, 2019, p. W2A.46.
[25] A. Salehiomran and Z. Jiang, "Fast BER distribution and neural networks for joint monitoring of linear and nonlinear noise-to-signal ratios," in *Proc. Opt. Fiber Commun. Conf.*, San Diego, California, 2020, p. M2J.3.
[26] A. S. Kashi, Q. Zhuge, J. C. Cartledge, S. A. Etemad, A. Borowiec, D. W. Charlton, C. Laperle and M. O'Sullivan, "Nonlinear signal-to-noise ratio estimation in coherent optical fiber transmission systems using artificial neural networks," *J. Light. Technol.*, vol. 36, no. 23, pp. 5424–5431, Dec. 2018.
[27] D. Lippiatt, S. Varughese, T. Richter, S. Tibuleac, and S. E. Ralph, "Joint linear and nonlinear noise estimation of optical links by exploiting carrier phase recovery," in *Proc. Opt. Fiber Commun. Conf.*, San Diego, California, 2020, p. Th2A.49.
[28] M. Cai, Q. Zhuge, H. Lun, M. Fu, L. Yi, and W. Hu, "Pilot-aided self-phase modulation noise monitoring based on artificial neural network," in *Proc. Asia Commun. Photonics Conf.*, p. 3, 2019.
[29] Y. Zheng, "Methodologies for cross-domain data fusion: an overview," *IEEE Trans. Big Data*, vol. 1, no. 1, pp. 16–34, Mar. 2015.
[30] H. Lun, X. Liu, M. Cai, M. Fu, Y. Wu, L. Yi, W. Hu, and Q. Zhuge, "Anomaly localization in optical transmissions based on receiver DSP and artificial neural network," in *Proc. Opt. Fiber Commun. Conf.*, San Diego, California, 2020, p. W1K.4.
[31] H. Lun, M. Fu, X. Liu, Y. Wu, L. Yi, W. Hu, and Q. Zhuge, "Soft failure identification for long-haul optical communication systems based on one-dimensional convolutional neural network," *J. Light. Technol.*, pp. 1–1, 2020.
[32] K. Simonyan and A. Zisserman, "Very deep convolutional networks for large-scale image recognition," *ArXiv14091556 Cs*, Apr. 2015, Accessed: Jul. 22, 2020. [Online]. Available: http://arxiv.org/abs/1409.1556.
[33] J. V. Davis, B. Kulis, P. Jain, S. Sra, and I. S. Dhillon, "Information-theoretic metric learning," in *Proc. Int. Conf. Mach. Learn.*, Oregon, 2007, pp. 209–216.
[34] F. Meng, K. Nikolovgenis, Y. Ou, R. Wang, Y. Bi, E. Hugues-Salas, R. Nejabati, and D. Simeonidou, "Field trial of gaussian process learning of function-agnostic channel performance under uncertainty," in *Proc. Opt. Fiber Commun. Conf.*, San Diego, California, 2018, p. W4F.5.
[35] I. de Miguel, R. J. Durán, T. Jiménez, N. Fernández, J. C. Aguado, R. M. Lorenzo, A. Caballero, I. T. Monroy, Y. Ye, A. Tymecki, I. Tomkos, M. Angelou, D. Klonidis, A. Francescon, D. Siracusa, and E. Salvadori, "Cognitive dynamic optical networks [Invited]," *J. Opt. Commun. Netw.*, vol. 5, no. 10, Art. no. 10, Oct. 2013.
[36] Z. Li, Y. Zhao, Y. Li, S. Rahman, X. Yu, and J. Zhang, "Demonstration of fault localization in optical networks based on knowledge graph and graph neural network," in *Proc. Opt. Fiber Commun. Conf.*, San Diego, California, 2020, p. Th1F.5.
[37] I. Sartzetakis, K. Christodoulopoulos, and E. Varvarigos, "Improving QoT estimation accuracy through active monitoring," in Proc. *Int. Conf. Trans. Opt. Net.*, Girona, Spain, Jul. 2017, pp. 1–4.
[38] H. B. McMahan, E. Moore, D. Ramage, S. Hampson, and B. A. y Arcas, "Communication-efficient learning of deep networks from decentralized data," *ArXiv160205629 Cs*, Feb. 2017, Accessed: Jun. 28, 2020. [Online]. Available: http://arxiv.org/abs/1602.05629.
[39] L. Velasco, B. Shariati, F. Boitier, P. Layec, and M. Ruiz, "Learning life cycle to speed up autonomic optical transmission and networking adoption," *J. Opt. Commun. Netw.*, vol. 11, no. 5, p. 226, May 2019.
[40] T. Panayiotou, G. Savva, I. Tomkos, and G. Ellinas, "Decentralizing machine-learning-based QoT estimation for sliceable optical networks," *J. Opt. Commun. Netw.*, vol. 12, no. 7, p. 146, Jul. 2020.

[41] Q. Zhuge, X. Zeng, H. Lun, M. Cai, X. Liu, L. Yi, and W. Hu, "Application of machine learning in fiber nonlinearity modeling and monitoring for elastic optical networks," *J. Light. Technol.*, vol. 37, no. 13, pp. 3055–3063, Jul. 2019.
[42] X. Liu, H. Lun, M. Fu, Y. Fan, L. Yi, W. Hu, and Q. Zhuge, "A three-stage training framework for customizing link models for optical networks," in *Proc. Opt. Fiber Commun. Conf.*, San Diego, California, 2020, p. Th3D.6.
[43] P. Poggiolini, "The GN model of non-linear propagation in uncompensated coherent optical systems," *J. Light. Technol.*, vol. 30, no. 24, pp. 3857–3879, Dec. 2012.
[44] O. V. Sinkin, R. Holzlohner, J. Zweck, and C. R. Menyuk, "Optimization of the split-step fourier method in modeling optical-fiber communications systems," *J. Light. Technol.*, vol. 21, no. 1, Art. no. 1, Jan. 2003.
[45] H. Li, Z. Xu, G. Taylor, C. Studer, and T. Goldstein, "Visualizing the loss landscape of neural nets," *ArXiv171209913 Cs Stat*, Nov. 2018, Accessed: Jun. 28, 2020. [Online]. Available: http://arxiv.org/abs/1712.09913.
[46] K. He, X. Zhang, S. Ren, and J. Sun, "Deep residual learning for image recognition," *ArXiv151203385 Cs*, Dec. 2015, Accessed: Jun. 28, 2020. [Online]. Available: http://arxiv.org/abs/1512.03385.
[47] O. Sagi and L. Rokach, "Ensemble learning: a survey," *Wiley Interdiscip. Rev. Data Min. Knowl. Discov.*, vol. 8, no. 4, Jul. 2018.
[48] R. E. Schapire, "The boosting approach to machine learning: an overview," in *Nonl. Est. Clas.*, vol. 171, NY: Springer New York, 2003, pp. 149–171.
[49] L. Breiman, "Bagging predictors," *Mach. Learn.*, vol. 24, no. 2, pp. 123–140, Aug. 1996.
[50] J. Sill, G. Takacs, L. Mackey, and D. Lin, "Feature-weighted linear stacking," *ArXiv09110460 Cs*, Nov. 2009, Accessed: Jun. 28, 2020. [Online]. Available: http://arxiv.org/abs/0911.0460.
[51] D. H. Wolpert, "Stacked generalization," *Neural Netw.*, vol. 5, no. 2, pp. 241–259, Jan. 1992.
[52] E. Bauer, "An empirical comparison of voting classification algorithms: bagging, boosting, and variants," *Machine learning*, 1999, 36.1-2: 105-139.
[53] T. Hastie, S. Rosset, J. Zhu, and H. Zou, "Multi-class AdaBoost," *Stat. Interface*, vol. 2, no. 3, pp. 349–360, 2009.
[54] T. Chen and C. Guestrin, "XGBoost: a scalable tree boosting system," *Proc. 22nd ACM SIGKDD Int. Conf. Knowl. Discov. Data Min.*, pp. 785–794, Aug. 2016.
[55] C. Zhang, D. Wang, C. Song, L. Wang, J. Song, L. Guan, and M. Zhang, "Interpretable learning algorithm based on XGBoost for fault prediction in optical network," in *Proc. Opt. Fiber Commun. Conf.*, San Diego, California, 2020, p. Th1F.3.
[56] Z. Zhou, J. Wu, and W. Tang, "Ensembling neural networks: many could be better than all," *Artif. Intell.*, vol. 137, no. 1–2, pp. 239–263, May 2002.
[57] D. Anguita, L. Ghelardoni, A. Ghio, L. Oneto, and S. Ridella, "The 'K' in K-fold cross validation," *Comput. Intell.*, p. 6, 2012.